\documentclass[prl,twocolumn,showpacs,amsmath,amssymb]{revtex4}
\usepackage{epsfig}
\usepackage{graphicx}
\usepackage{dcolumn}
\usepackage{bm}
\begin{document}
\title{Probing a nonequilibrium Einstein Relation in an aging colloidal glass}

\author{B\'ereng\`ere Abou and Fran\c cois Gallet}
\affiliation{Laboratoire de Biorh\'eologie et
Hydrodynamique Physico-chimique UMR CNRS 7057 and F\'ed\'eration de
Recherche Mati\`eres et Syst\`emes Complexes FR 2438,\\
2 Place Jussieu,
Case 7056,\\
75251 Paris Cedex 05, FRANCE}

\date{\today}

\begin{abstract}
We present a direct experimental measurement of an effective
temperature in a colloidal glass of Laponite, using a micrometric bead
as a thermometer. The nonequilibrium fluctuation-dissipation relation,
in the particular form of a modified Einstein relation, is
investigated with diffusion and mobility measurements of the bead embedded in the glass. We observe an
unusual non-monotonic behavior of the effective temperature~: starting from the bath
temperature, it is found to increase up to a maximum value, and then
decreases back, as the system ages. We show that the observed
deviation from the Einstein relation is related to the relaxation
times previously measured in dynamic light scattering
experiments.
\end{abstract}

\maketitle

At thermodynamical equilibrium, the fluctuation-dissipation theorem (FDT)
relates the response function
of the system to its time autocorrelation function. This
theorem may display several forms, depending on the considered observable (Nyquist formula, Einstein
relation), but involves a single thermodynamic parameter, which is the
equilibrium temperature. However, FDT applies only to ergodic systems
at equilibrium, and is not expected to hold for non-equilibrium systems, like
glasses and gels, which exhibit relaxation times
longer than or comparable to the observation timescale.

Recently, many efforts have been devoted to apply statistical physics
concepts to such out-of-equilibrium systems. In particular,
fluctuation-dissipation relations have been extended with the help of
a timescale dependent effective temperature, different from the bath
temperature, and that has been shown to display many of the properties
of a thermodynamic temperature \cite{cug93,cug97}. This deviation from
equilibrium FDT has been observed in many numerical simulations
\cite{cug93, parisi, barrat, sellitto, marinari,barrat2, barrat3,
makse,fielding}. In these models and simulations, violations are
expected to occur when the characteristic observation time $1/\omega $
is of the same order or greater than the aging time $t_w$. On the
other hand, there are only few experimental studies of FDT violations
in aging materials \cite{grigera,bellon,herisson,danna,oja}. In all
experimental systems, violations are observed when $1/\omega $ is
smaller than $t_w$. This suggests that the aging time $t_w$ is not the
only relevant parameter to describe aging, as usually considered in
numerical simulations, and that microscopic processes, charaterized by
the distribution of relaxation times in the system, must be
considered.

In this letter we measure the evolution with aging time of an
effective temperature $T_{\mbox{\scriptsize eff}}$ in a colloidal glass of Laponite. This
is achieved by simultaneously measuring
the mobility -- using an optical tweezer -- and the position fluctuations
of a micrometric bead
embedded in the glass. We find that, starting from the bath
temperature, $T_{\mbox{\scriptsize eff}}$ increases with $t_w$, up to 1.8 times the bath
temperature, and decreases back upon further aging. We propose a
consistent interpretation of this unusual behavior by considering the
evolution of the relaxation times distribution in the glass.

Let us consider a diffusing particle of mass $m$ evolving in a
stationary medium. Its motion is described by a generalized Langevin
equation $m~dv/dt = F(t) -m \int_{-\infty}^{+\infty} \gamma(t-t')
~v(t') ~dt' $, in which $v(t) = dx/dt$ is the particle
velocity, $F(t)$ the random force acting on the particle and
$\gamma (t)$ a delayed friction kernel that takes into account the
viscoelastic properties of the medium. If the surrounding stationary medium is in thermal equilibrium at temperature $T$,
one can derive a generalized Einstein relation, which is a specific
form of FDT, $s^2 \langle \Delta \hat{x}^2
(s) \rangle = 2 kT \hat{\mu} (s) $, in which $\langle \Delta \hat{x}^2(s) \rangle $ is the
mean-square displacement Laplace transform and $\hat{\mu}(s)=1 / m
\hat{\gamma} (s)$ the mobility Laplace transform, if inertia
is neglected.

The general situation of a particle diffusing in an
out-of-equilibrium environment is much more difficult to
describe. This point has been developed in details in
\cite{pottier} using a generalized
Langevin formalism. The FDT is extended with the help of
a frequency-dependent effective temperature $T_{\mbox{\scriptsize eff}}
(\omega)$, parametrized by the age $t_w$ of the
system, and which satisfies \cite{npottier} : 
\begin{equation}
T_{\mbox{\scriptsize eff}} (\omega) \Re e {\mu(\omega)} = \Re e [ \mu (\omega)
\Theta(\omega) ] 
\end{equation}
where 
\begin{equation}
s^2 \langle \Delta \hat{x}^2 (s) \rangle = 2 k \hat{\Theta} (s) ~\hat{\mu} (s)
\end{equation}
The two equations (1) and (2) define an out-of-equilibrium Einstein relation, for a
given aging time $t_w$. The parameter $\hat{\Theta} (s)$ is related to $\Theta
(\omega)=\hat{\Theta}(s = -i \omega)$. 

The principle of our experiment to measure the effective temperature
in an out-of-equilibrium system is the following : micrometric probes
are immersed in a colloidal glass to allow the measurements, at the
same aging time $t_w$, of both their fluctuating position and
mobility. $T_{\mbox{\scriptsize eff}}$ is then obtained from equations
(1) and (2). The colloidal glass consists in an aqueous suspension of
Laponite RD, a hectorite synthetic clay provided by Laporte
Industry. The preparation procedure of the glass has been addressed in
details in \cite{jorlapo}. These aqueous suspensions form glasses for
low volume fraction in particles \cite{wigner}. Being in a ``liquid''
state right after preparation, the suspension becomes more and more
viscoelastic with time. Since the physical properties of the
suspension depend on the time $t_w$ elapsed since preparation, the
sample is said to age. Aging can be seen through the evolution of both
the viscoelastic properties and of the colloidal disks diffusion
\cite{jorlapo,aboupre64}. This glass, obtained at the ambient
temperature, is optically transparent. Moreover, it presents other
advantages that make possible to measure an effective temperature with
tracer beads. First, the preparation procedure allows to obtain a
reproducible initial state, leading to an accurate determination of
the origin $t_w = 0$. Second, the Laponite suspensions age on
timescales that depend on the particles concentration. We are thus
able to choose the aging timescales of the glass by adjusting this concentration. With a volume fraction of $2.3 \% $wt, the
glass evolves slowly enough to allow two successive measurements -
fluctuation and dissipation - without significative aging of the
sample. The two measurements are thus performed at the same $t_w$.

The experiments were carried out in a square chamber -- 20 x 20
mm$^2 $ -- made of a microscope plate and a coverslip separated by a thin
spacer ($0.1$ mm thickness). The beads are suspended in the glass right after its preparation. The chamber is then filled with the
suspension, sealed with vacuum grease and mounted on a piezoelectric
stage on the plate of an optical microscope. The probes are latex and
silica beads, in very low concentration (respectively $10^{-4} \%$ and
$4.10^{-4}  \%$ in volume). Latex beads ($1.0 \pm 0.1 \mu$m in diameter,
Polysciences, Inc.), were preferentially used for fluctuation
measurements : since they do not deposit during the experiment, their
random motion is not perturbated by the chamber walls. Silica beads
($2.1 \pm 0.1 \mu$m in diameter, Bangs Lab Inc.) were used for dissipation
measurements, because they are more efficiently trapped by the optical
tweezers. The diameters of the two kinds of probes are close to each
other, thus the comparison between the results of the fluctuation and
dissipation measurements, once rescaled to the same diameter, is
meaningful.

Let us first focus on the two-dimensional brownian
motion of a tracer bead immersed in the glass. At a given aging time
$t_w$, we record the fluctuating motion of a 1 $\mu$m latex bead during 8
s, with a fast CCD camera sampling at 125 Hz (Kodak, PS-220). A
digital image analysis allows to track the bead positions $x(t)$ and
$y(t)$ close to the focus plane of the microscope objective. For each
bead, we calculate the time-averaged mean-square displacement $
\langle \Delta r^2 (t) \rangle _{t'} = \langle( x(t'+ t) - x(t') )^2 + (y(t' + t) - y(t')
)^2 \rangle_{t'} = 2 \langle \Delta x^2 (t) \rangle_{t'}$. To preserve a good statistics,
we keep the data of $\langle\Delta r^2 (t)\rangle_{t'}$ in the range $0.008 < t < 1$ s. The
glass remains in a quasi-stationary state during the recording, which
takes a short time compared to the aging timescale. The quantity
$\langle\Delta r^2 (t)\rangle_{t'}$ can thus be identified to the
ensemble-averaged mean-square displacement $\langle \Delta r^2
(t)\rangle_E$. Fig. 1 shows the mean-square displacement of latex beads
immersed in the colloidal glass, as a function of time t, for various
aging times $t_w$. The mean-square displacement is well described by a
power-law behavior $ \langle \Delta r^2 (t)\rangle_E = D t^{\alpha}$ over the full
range $0.008 < t < 1 $ s. Upon increasing on $t_w$,
the exponent decreases from $1.05 \pm 0.05$ at the earliest aging times
to $0.25 \pm 0.05$ at long aging times. This indicates a nearly
diffusive behavior of the tracer bead at short aging times, that
becomes sub-diffusive as the glass ages. 
\begin{figure}[b]
\includegraphics[angle=-90,width=5.0cm]{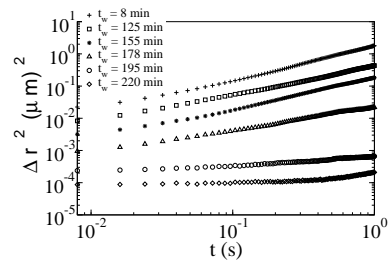}
\caption{ \label{meansquare} Mean-square displacement of a 1 $\mu$m
latex bead immersed in the glass, as a function of
time. The curves correspond to different $t_w =
8, 125, 155, 178, 195$ and $220$ minutes from top to bottom. The
fluctuating motion is purely diffusive at short $t_w$ and
becomes sub-diffusive as the glass ages. }
\end{figure}
We describe now the
measurement of the mobility $\mu(\omega)$, at a given frequency
$\omega $ for various aging times. We recall that this measurement is
performed right after the fluctuation motion recording, at the same
aging time $t_w$. As the Laponite suspension is a viscoelastic fluid,
the bead mobility $\mu(\omega) = |\mu(\omega)| e^{i \phi(\omega)} $ is
a complex number. We thus need to measure the phase and modulus of the
tracer mobility. We use an optical tweezer to trap a 2.1 $\mu$m silica
bead immersed in the glass. Trapping is achieved by focusing a
powerful infrared laser beam (Nd YAG, Spectra-Physics, $P_{max} = 600$
mW) through a microscope objective of large numerical aperture \cite{henon}. The
trapping force $F$ on a small dielectric object like a silica bead is
proportional to the intensity gradient in the focusing region. To
first order, one can write $F = -kx$, where $x$ represents the
distance of the trapped object from the trap center. The stiffness $k$
is known from an independent
calibration. Once the bead is trapped, we make the experimental
chamber oscillate by monitoring the displacement $x_p \exp(i \omega
t)$ of a piezoelectric stage. As a consequence, the viscoelastic glass
exerts a sinusoïdal force $F' \exp(i \omega t)$ on the bead. We record
with the fast camera the bead movement, and measure by conventional
image analysis its displacement $x \exp{(i \omega t)}$ from the trap center. At a
given frequency $\omega$, the force $F'(\omega)$ is given by
$F'(\omega) = v(\omega)/ \mu (\omega)$, where $v(\omega) = i \omega (x_p -
x)$ is the relative glass / bead velocity, and $\mu(\omega)$ the
Fourier transform of the bead mobility. In our range of experimental
frequencies ($0.5 < f < 10$ Hz), the bead inertia is negligible, so
that we can simply use the relation $F + F' = 0$ to calculate $|\mu(\omega)|$
and $\phi(\omega)$. Notice that the piezoelectric stage and the camera are
triggered by two synchronized signals, numerically generated by a PC
computer, so that the phase shift between the force and the bead
movement can be accurately measured. Fig. 2 (a,b) show the complex
mobility of the tracer bead as a function of the aging time $t_w$ for
various frequencies of the applied force. 
\begin{figure}[b]
\includegraphics[width=5.5cm]{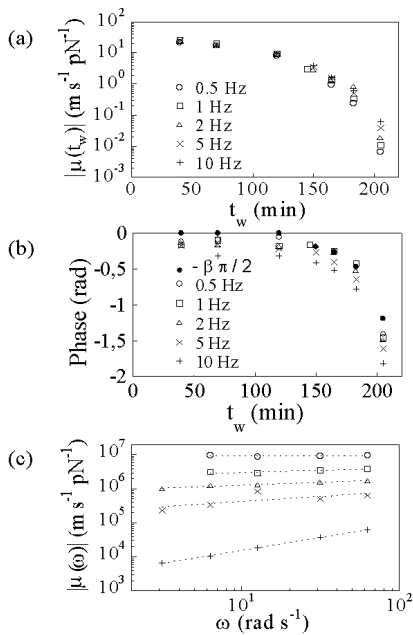}
\caption{ \label{mobility} Modulus (a) and phase (b) of the complex
mobility $\mu(\omega) = |\mu(\omega)| e^{i \phi(\omega)} $ of the
tracer bead as a function of $t_w$, for various frequencies of the
applied force. The full circles in (b) correspond to $\phi= -
\beta(t_w) \pi/2$; (c) : mobility modulus versus $\omega$ for various
$t_w$; from top to bottom, $t_w = 120, 145, 165, 183, 205$
min. At low $t_w$, $|\mu(\omega)|$ is independent of $\omega$. Upon increasing $t_w$, the modulus is well fitted by a
power-law $|\mu(\omega)|=\mu_0 \omega^{\beta}$ with $\beta$ only
depending on $t_w$. }
\end{figure}

Two more steps are necessary to calculate the effective temperature
from both fluctuation and mobility data. First, the mean-square
displacement is numerically Laplace transformed to the frequency
domain $0.15 - 20$ Hz. In this range, we find that the Laplace
transform is well adjusted by a power-law $\langle \Delta \hat{r}^2
(s) \rangle = a s^{- b}$. As expected from the observation $\langle
\Delta r^2 (t)\rangle _E = D t^{\alpha}$, the relation $b= \alpha + 1$
is accurately verified. Second, we analyze the frequency dependence of
the mobility. In the experimental frequency range, $|\mu(\omega)|$ is
well fitted by a power law $|\mu(\omega)| = \mu_0 \omega^{\beta}$, as
shown in Fig. 2c. The exponent $\beta$ increases with $t_w$ from zero
at low aging times to $0.75 \pm 0.05$ at the end of experiment. To a
first-order approximation, we consider that the phase $\phi$ is
independent of the frequency. Within a good approximation, its
dependence on $t_w$ can be related to $\beta$ by $\phi= - \beta(t_w)
\pi/2$ as shown in Fig. 2b. From the analytical form $\mu(\omega) =
\mu_0 \omega^{\beta} \exp{(-i \beta \pi /2)}$, the Laplace transform
$\hat{\mu}(s) = \mu(\omega= is)=\mu_0 s^\beta$ is derived by
analytical continuation. Using equation (2), we calculate the function
$\hat{\Theta}(s)$, parametrized by $t_w$, and derive $\Theta (\omega)=
\hat{\Theta}(s= - i\omega)$. Finally, the effective temperature, at a
given $t_w$, is :
\begin{equation}
k T_{\mbox{\scriptsize eff}}
(\omega)=\frac{a}{4\mu_0} \frac{cos((b-2)\pi/2)}{cos(\beta \pi/2)} \omega^{2-b-\beta}
\end{equation}

The dependence of $T_{\mbox{\scriptsize eff}}$ on $t_w$ is shown in Fig. 3, at a frequency $f =
1$ Hz. The results have been averaged over three realizations. At the
earliest $t_w$, the effective temperature is close to the bath
temperature $T_{\mbox{\scriptsize bath}} = 300$ K. Upon increase on
$t_w$, $T_{\mbox{\scriptsize eff}}$ increases up to 1.8 times the bath
temperature and then decreases towards $T_{\mbox{\scriptsize bath}}$
upon further increase on $t_w$. This is the first time that such a
behavior -- increase of $T_{\mbox{\scriptsize eff}}$ followed by a
decrease -- is observed in a glassy system. A realistic explanation of
this non-monotonic behavior is proposed in the next
paragraph. Moreover, our results are in contradiction with electrical
measurements of FDT performed in the same system \cite{bellon}, where
$T_{\mbox{\scriptsize eff}}$ is found to decrease with $t_w$ and
$\omega$ and is larger by about one order of magnitude. A possible
origin of this discrepancy may be the choice of different observables
in the two experiments \cite{fielding}.
\begin{figure}[b]
\includegraphics[angle=-90,width=5.0cm]{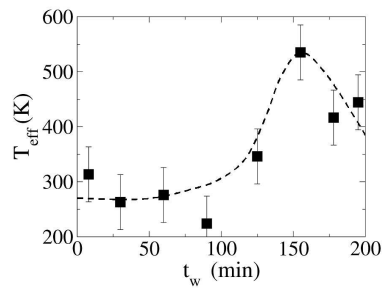}
\caption{ \label{temp} Effective temperature of the colloidal glass versus $ t_w$,
measured at a frequency $f = 1$ Hz. Upon increase on $t_w$, $T_{\mbox{\scriptsize eff}}$ increases
up to 1.8 times the bath temperature and decreases back to $T_{\mbox{\scriptsize bath}}$. The dashed line is a guide for the eyes.   }
\end{figure}

An interpretation of the
dependence of $T_{\mbox{\scriptsize eff}}$ on $t_w$ is provided by Dynamic Light Scattering (DLS)
and Diffusive Wave Spectroscopy (DWS) experiments, previously
performed in colloidal glasses of Laponite \cite{aboupre64,lequeux}. The resulting
distribution function of relaxation times $ P(\tau)$ is schemed in
Fig. 4. Upon increasing $t_w$, part of the modes distribution function
shift to larger times, while the mode at $\tau \sim 0.1 $ ms remains
unchanged. When probing our experimental system at a typical frequency
$f = 1$ Hz, three situations occur. At the earliest $t_w$, the fast modes $\tau 
<< 1/ f$ allow the system to thermalize with the bath (Fig. 4a). The
measured effective temperature is the bath temperature. Upon increase
on $t_w$, modes $\tau \sim  1/ f$ appear in the system as seen in Fig. 4b. On this
observation timescale $1/f$, the system is out-of-equilibrium~: the
measured temperature becomes different from the bath temperature and
is timescale dependent. Deviation from the equilibrium Einstein relation is thus
observed. Finally, for very long $t_w$, the fast modes allow the
system to thermalize with the bath while the slow ones $\tau  \gg 1/ f$ do not
play any role at the experimental timescale. $T_{\mbox{\scriptsize eff}}$ is then expected to
reduce back to $T_{\mbox{\scriptsize bath}}$. These different situations are clearly
identified in Fig. 3. However the experimental set-up does not allow
to observe a further decrease of $T_{\mbox{\scriptsize eff}}$ at long
$t_w$. Indeed, beyond $t_w = 205$
minutes, the mobility modulus becomes smaller than $10^{-2}$ m s$^{-1 }$pN$^{-1 }$. In
this range, the optical tweezer is not powerful enough to induce a
detectable motion of the bead. 

In these experiments, deviations are observed when $\omega
\tau \sim 1$. This suggests that, besides the waiting time $t_w$, the
distribution of relaxation times must be included in models to
achieve an accurate description of aging.

\begin{figure}[b]
\includegraphics[width=6.0cm]{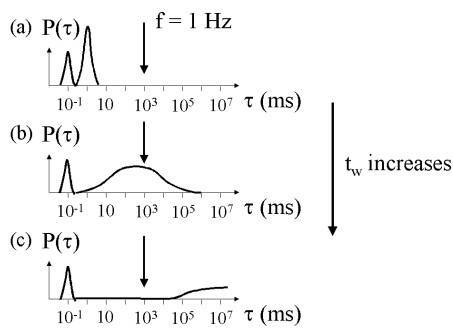}
\caption{ \label{tau} Scheme of
the distribution function of relaxation times $P(\tau)$ in the glass of Laponite (typically $2.5 \%$ wt) at different $t_w$. Upon
increasing $t_w$, part of the modes distribution shifts towards larger
times, while the mode at $\tau \sim 0.1$ ms remains unchanged. The arrow
represents the measurements timescale $1/f$. }
\end{figure}

Let us now comment the dependence of
$T_{\mbox{\scriptsize eff}}$ on the frequency. Since we experimentally find that $2 - b - \beta > 0$
at all $t_w$, $T_{\mbox{\scriptsize eff}}$ increases with $\omega$ as $\omega^{2 - b - \beta}$, according to equation
(3). This behavior is consistent with our interpretation at least in
the first situation. However, in our picture, one expects $T_{\mbox{\scriptsize eff}}$ to
decrease with $\omega$ at long $t_w$. Actually, one can see from Fig. 2b that $\phi$
can no longer be accurately considered as independent of $\omega$ at long
$t_w$. Thus, the analytical interpolation $\mu(\omega) = \mu_0 \omega^{\beta} \exp{-i \beta \pi
/2}$ probably breaks down. This difficulty will be soon overcomed by a
direct measurement of $\hat{\mu}(s)$ from a creep experiment. 

This work provides a test of the nonequilibrium Einstein relation in a
colloidal glass, using diffusion and mobility measurements on a
micrometric probe. We observe for the first time in a glassy system a
non-monotonic behavior of $T_{\mbox{\scriptsize eff}}$ with $t_w$. It seems likely that this behavior is directly related to the
evolution of the relaxation times distribution. Notice that deviations from the Einstein relation, found in
supercooled liquids even at equilibrium, were explained by spatial
heterogeneities \cite{ediger}. Future works will have to analyze how such heterogeneities interplay with the concept of
$T_{\mbox{\scriptsize eff}}$ -- which is an ensemble-averaged quantity. Another open question is
whether this effective temperature has a real thermodynamics meaning
\cite{cug97, fielding}. The answer will come from experimental tests
of FDT involving other physical
observables. 

\begin{acknowledgments}
We are indebted to N. Pottier for enlightening
discussions and comments. We thank P. Monceau for numerical Laplace
transformations, M. Balland for help in programmation, and
L. Cugliandolo for fruitful exchanges.  
\end{acknowledgments}


\begin{thebibliography}{99}

\bibitem{cug93} Cugliandolo, L., and Kurchan, J. {\it
Phys. Rev. Lett.} {\bf 71}, 173 -176 (1993).
 
\bibitem{cug97}
Cugliandolo, L.F., Kurchan, J. and Peliti, L. {\it Phys. Rev. E} {\bf 55}, 3898 - 3914 (1997).  


\bibitem{parisi}
Parisi, G. {\it Phys. Rev. Lett.} {\bf 79}, 3660 - 3663 (1997).

\bibitem{barrat}
Barrat, A. {\it Phys. Rev. E } {\bf 57}, 3629 - 3632 (1998).   

\bibitem{sellitto}
Sellitto,
M. {\it Eur. Phys. J. B.} {\bf 4}, 135 - 138 (1998).

\bibitem{marinari}
Marinari, E.,
Parisi, G., Ricci-Tersenghi, F., and Ruiz-Lorenzo, J.J. {\it J. Phys. A : Math Gen.} {\bf 31}, 2611 - 2620 (1998).

\bibitem{barrat2}
 Barrat,
J.-L., and Kob, W. {\it Europhys. Lett.} {\bf 46}, 637 - 642 (1999).

\bibitem{barrat3}
Berthier, L., Barrat, J.-L., and Kurchan. J. {\it Phys. Rev. E} {\bf 61}, 5464 -
5472 (2000).

\bibitem{makse}
Makse, H. A., and Kurchan, J. {\it Nature} {\bf 415}, 614 - 617 (2002).  

\bibitem{fielding}
Fielding,
S. and Sollich, P. {\it Phys. Rev. Lett.} {\bf 88}, 050603-1 -
050603-4 (2002).
 
\bibitem{grigera} Grigera, T. S. and Israeloff, N. E. {\it Phys. Rev. Lett.} {\bf 83}, 5038 - 5041 (1999).
 

\bibitem{bellon}
Bellon, L., Ciliberto, S. and Laroche, C. {\it Europhys. Lett.} {\bf 53}, 511 - 517 (2001).
 
\bibitem{herisson} H\'erisson, D. and Ocio, M. {\it Phys. Rev. Lett.} {\bf 88}, 257202
- 257205 (2002).
 
\bibitem{danna}
D'Anna, G., Mayor, P., Barrat, A., Loreto, V. and Nori, F. {\it Nature} {\bf 424},
909 - 912 (2003). 

\bibitem{oja}
Ojha., R. P., Lemieux, P.-A., Dixon, P. K.,
Liu, A. J. and Durian, D. J. {\it Nature} {\bf 427}, 521 - 523 (2004). 

\bibitem{pottier} Pottier, N. {\it Physica A} {\bf 317}, 371 - 1382 (2003);
Pottier, N. and Mauger, A. {\it Physica A} {\bf 332}, 15 - 28 (2004).

\bibitem{npottier}  Pottier, N. cond-mat/0404613, to appear in {\it Physica
  A}. 

\bibitem{jorlapo} Abou, B., Bonn, D., and Meunier, J. {\it
J. Rheol.} {\bf 47}, 979 - 988 (2003).


\bibitem{wigner} Bonn, D., Tanaka, S., Wegdam, G. H., Kellay, H., and
Meunier, J. {\it Europhys. Lett.}
{\bf 45}, 52 - 57 1(1999).

\bibitem{aboupre64}
Abou, B., Bonn, D.,
and Meunier, J. {\it Phys. Rev. E} {\bf 64},
021510 - 021513 (2001). 

\bibitem{henon}
H\'enon, S., Lenormand, G., Richert, A., and Gallet, F. {\it Biophys. J. } {\bf 76}, 1145 - 1151 (1999).

 
\bibitem{lequeux}
Knaebel A., Bellour M., Munch J.-P.,
Viasnoff V., Lequeux, F. and Harden, J. L. {\it Europhys. Lett.} {\bf 52}, 73 - 79 (2000). 

\bibitem{ediger}
Swallen, S. F., Bonvallet, P. A., McMahon, R. J., and Ediger, M. D. {\it Phys. Rev. Lett.} {\bf 90}, 015901 1-4 (2003).

\end{thebibliography}
\end{document}